\documentclass[twocolumn,showpacs,amsmath,amssymb,superscriptaddress]{revtex4}

\usepackage{graphicx}
\usepackage{dcolumn}
\usepackage{bm}
\usepackage{epsfig}
\usepackage{slashed}

\begin{document}

\def\0#1#2{\frac{#1}{#2}}
\def\bct{\begin{center}} \def\ect{\end{center}}
\def\beq{\begin{equation}} \def\eeq{\end{equation}}
\def\bea{\begin{eqnarray}} \def\eea{\end{eqnarray}}
\def\nnu{\nonumber}
\def\n{\noindent} \def\pl{\partial}
\def\g{\gamma}  \def\O{\Omega} \def\e{\varepsilon} \def\o{\omega}
\def\s{\sigma}  \def\b{\beta} \def\p{\psi} \def\r{\rho}
\def\G{\Gamma} \def\k{\kappa} \def\l{\lambda} \def\d{\delta}

\title{A practical method in calculating one loop quantum fluctuations to the energy of the non-topological soliton}
\author{Song~Shu}
\affiliation{Department of Physics and Electronic Science, Hubei
University, Wuhan 430062, China} \affiliation{Department of
Physics and Astronomy, Stony Brook University, Stony Brook NY 11794
USA}
\begin{abstract}
I have used a practical method to calculate the one-loop quantum
correction to the energy of the non-topological soliton in
Friedberg-Lee model. The quantum effects which come from the
quarks of the Dirac sea scattering with the soliton bag are
calculated by a summation of the discrete and continuum energy
spectrum of the Dirac equation in the background field of soliton.
The phase shift of the continuum spectrum is numerically
calculated in an efficient way and all the divergences are removed
by the same renormalization procedure.
\end{abstract}
 \maketitle

Non-topological soliton models which are effective models inspired
from the underlying QCD theory are phenomenologically successful
in describing the low energy nuclear physics. However, the main
calculation methods in these models are based on mean field
approximation, in other words treating the fields
classically~\cite{ref1,ref2,ref3,ref4}. The quantum corrections in
the background fields of spatially non-trivial configurations are
very difficult to calculate. This is partly due to the fact that
these calculations are nonlocal. During the past decades different
calculation methods and approximate schemes have been developed on
this problem~\cite{ref5,ref6,ref7,ref8,ref9,ref10,ref11,ref12}. As the calculation of quantum corrections of solitons is much more complex than those usual calculations of quantum loop corrections of trivial background fields, most studies on this problem are based on the derivative expansion method~\cite{ref5,ref6,ref7,ref8,ref9}. The renormalization in this method is a very nontrivial task.
One remarkable calculation method was that developed by Farhi, Graham, Haagensen and Jaffe~\cite{ref13}. It is a systematic and efficient scheme for
calculating the quantum corrections about static field
configuration in renormalizable field theories, in which all the
divergences are removed by the same renormalization
procedure. As originally this method was applied in the Higggs like models and the main interest was focused on studying solitons in the standard electroweak models~\cite{ref13,ref14}, there are no applications of this method, as far as I know, in strong interaction hadronic models, like the Friedberg-Lee(FL) model, the linear sigma model and other QCD effective models. In recent years topological solitons in strong interaction QCD theory have drawn lots of attentions~\cite{ref15,ref16}. One needs an efficient method to calculate the quantum correction of the soliton in effective QCD theories~\cite{ref17}. So in this paper as the first small step I want to introduce this method to calculate the one loop quantum fluctuation of the non-topological soliton in the FL model. In this method
one makes the energy level summation by calculating the discrete
and continuous energy spectrum and the continuum contribution is
determined through evaluating scattering phase shift in a concise
way. The renormalization of the field configuration energy could
be done in a manner consistent with on-shell mass and coupling
constant renormalization in the perturbative sector. Comparing to
the precedent calculation technics in the literatures this method
is more efficient and practical.

Consider the Lagrangian of the FL model, \bea {\cal
L}=\bar\psi(i\gamma_\mu\pl^\mu-g\s)\psi+\012(\pl_\mu\s)(\pl^\mu\s)-U(\s),
\eea where\bea U(\s)=\01{2!}a\s^2+\01{3!}b\s^3+\01{4!}c\s^4+B.
\eea $\p$ represents the quark field, and $\s$ denotes the
phenomenological scalar field. $a, b, c$, and $g$ are the
constants which are generally fitted in producing the properties
of hadrons properly. $B$ is the bag constant. In the background of a nontrivial $\s$ field there
might be some bound state levels with energy $0<E_n<m$ which can
allocate the quarks lowering the energy of the whole system at the
expense of creating such nontrivial configuration of $\s$ field. This is the non-topological soliton solution in the FL model.
Generally speaking the spherical configuration of the $\s$ field
will take the following form, \beq \s(r)=\s_v
-\0{\s_0}{1+e^{(r-R)/r_0}}, \label{bag}\eeq where the second term
on the r.h.s of the equation is a Woods-Saxon potential well with
depth $\s_0$. Inside a sphere of radius $R$ the $\s$ field almost
vanish, while outside the well it takes its asymptotic vacuum
value $\s_v$. The valence quarks are bounded in the well and form a classical soliton.
By fitting the hadron properties the model parameters could be
fixed but not uniquely. There are some flexibilities in choosing
the parameters. For baryons if one takes $N=3$ and chooses
one set of values of parameters as $a=17.7fm^{-2}, b=-1457.4fm^{-1},
c=20000, g=12.16$~\cite{ref2}, one obtains a classical soliton energy $E_{cl}\approx 6.4
fm^{-1}\approx 1262MeV$.

Next I will study the quantum correction of the classical soliton. In principle the quantum corrections in FL model should
include loop corrections from both quark fields and the $\s$
field. However since the $\s$ field is only a phenomenological
field describing the long-range collective effects of QCD, the
loop corrections coming from the sigma field will be ignored. So I
just consider one loop fluctuations from the quark field in a
static nontrivial configuration of $\s$ field background. In this case
the one loop effective action after integrating out the quark
field is given by \beq S_{eff}=S_{cl}+S_{ct}-i\log \det D \eeq
where $S_{cl}$ is the classical part, $S_{ct}$ is the counterterm
part and $D$ is the Dirac operator which general form is
$D=i\g_{\mu}\pl^{\mu}-g\s(r)$. The total energy could be derived
by $E_{tot}=-S_{eff}/\int dt$ and the result is \beq
E_{tot}=E_{cl}+E_{ct}+E_{vac}^{\psi}, \eeq where $E_{ct}$ is the
necessary renormalization counterterm and $E_{vac}^{\psi}$ is the vacuum correction as a
result of the energy level summation from both discrete and
continuum spectrum. The whole energy spectrum is determined by the
following stationary dirac equation \beq
[-i\alpha\cdot\vec\nabla+\beta g\s(r)]\psi=E\psi. \label{dirac}\eeq One could solve the Dirac equation and get the
continuous energy spectrum $E(k)=\sqrt{k^2+m^2}$ where $m=g\s_v$ and some possible
discrete energy spectrum $0<E_n<m$, thus the energy level sum over
discrete and continuous spectrum is \beq
E_{vac}^{\psi}=-\sum\limits_{n}E_n-\sum\limits_{l}(2l+1)\int
dk\r_l(k)E(k), \label{evac} \eeq where $\r_l(k)$ is the density of
states in momentum space with the angular momentum quantum number
$l$ and $(2l+1)$ is the degenerate factor of the angular momentum
projection. The density of states $\r_l(k)$ will relate to the scattering
phase shift $\d_l(k)$ in the following way~\cite{ref13} \beq
\r_l(k)=\r_l^{\textmd{free}}(k)+\01\pi\0{d\d_l(k)}{dk}, \eeq where
$\r_l^{\textmd{free}}(k)$ is the density of states when the background $\s$ field
is trivial. In our case this part will be subtracted from
the density of states since I only consider the quantum
corrections of the nontrivial background $\s$ field.

The main difficulties come from the calculations of the scattering
phase shift $\d_l(k)$ and the renormalization. To eliminate the
divergence of the integral over continuum spectrum in equation
(\ref{evac}) the phase shift needs to be rendered by a Born
approximation according to the stand method in quantum mechanics.
In one loop calculation only the first and second Born
approximation should be subtracted from the phase shift. Therefore
the subtracted phase shift is defined as \beq \bar\d_l(k)\equiv
\d_l(k)-\d_l^{(1)}(k)-\d_l^{(2)}(k), \label{delta} \eeq in which
$\d_l^{(1)}(k)$ and $\d_l^{(2)}(k)$ are the first and second Born
approximations to $\d_l(k)$. These phase shifts can be determined by solving the equation (\ref{dirac}). In order to solve it
one need to decompose the quark field into \bea
\psi(\vec r)=\01r\left(
\begin{array}{c}
F(r)\\
i\vec\s\cdot\hat{\vec{r}} G(r)
\end{array}\right)y_{\kappa m},
\eea where $y_{\kappa m}\equiv y_{jm}^{l}$ is the two-component
Pauli spinor harmonic, $\kappa$ is the Dirac quantum number $\kappa=-(l+1)$ and $\hat{\vec r}$ is the spatial unit vector. Substitute it into the equation (\ref{dirac}) one obtains two coupled first order radial equations of upper component $F$ and lower component $G$. These two equations can be decoupled to two second order differential equations about $F$ and $G$. One can use either of them to evaluate the phase shift. The equation of upper component $F$ is \bea F''&-&\0{g\s'}{E+g\s}F'-\left[\0\k
r\0{g\s'}{E+g\s}\right. \nnu \\ &+&\left.\0{\k(\k+1)}{r^2}-(E^2-g^2\s^2)\right]F=0,
\label{F} \eea where the prime denotes the differentiation with
respect to $r$. In the following I
will use this equation to calculate the phase
shift. When $r>>R$ the asymptotic form of equation (\ref{F}) is
\beq F''-\left[\0{\k(\k+1)}{r^2}-k^2\right]F=0, \eeq where
$k^2=E^2-g^2\s^2_v$. The solutions will be spherical Hankel
functions. At the same time for equation (\ref{F}) the solution
should satisfy that $F(r)\to 0$ as $r\to 0$. Thus one could
introduce two linearly independent solutions to equation (\ref{F})
as \beq F^{(1)}_l(r)=e^{i\b_l(k,r)}rh^{(1)}_l(kr), \eeq \beq
F^{(2)}_l(r)=e^{-i\b^*_l(k,r)}rh^{(2)}_l(kr), \eeq where
$h^{(1)}_l(kr)$ and $h^{(2)}_l(kr)$ are the Hankel functions of
the first and second kinds and $h^{(2)}_l(kr)=h^{(1)*}_l(kr)$. The
function $\b_l(k,r)$ should satisfy $\b_l(k,r)\to 0$ as $r\to
\infty$. Then the scattering solution is \beq
F_l(r)=F^{(2)}_l(r)+e^{i\d_l(k)}F^{(1)}_l(r), \eeq and obeys
$F_l(0)=0$, which leads to the result of the scattering phase
shift \beq \d_l(k)=-2\textmd{Re}\b_l(k,0), \eeq where
$\textmd{Re}$ means the real part. By substituting $F^{(1)}_l$
into equation (\ref{F}) one could obtain the equation of $\b_l$
\bea &&i\b''_lrh_l+2i\b'_l(h_l+rh'_l)-\b^{\prime
2}_lrh_l-\0{g\s'}{E+g\s}(i\b'_lrh_l+h_l\nnu \\ &&+rh'_l)-\left[\0{\k}{r}\0{g\s'}{E+g\s}+g^2(\s^2-\s^2_v)\right]rh_l=0.
\label{ricatti} \eea In the fixed background soliton field of
$\s(r)$ this equation could be numerically solved to obtain the
phase shift $\d_l(k)$.

To get the Born approximation to the phase shift one should expand
$\b_l$ in powers of $g$ as \beq \b_l=g\b_{l1}+g^2\b_{l2}+\cdots.
\label{beta} \eeq Substituting the expansion (\ref{beta}) into
equation (\ref{ricatti}) and neglecting the higher order terms
$O(g^3)$ one can obtain a set of coupled differential equations
about $\b_{l1}$ and $\b_{l2}$ as \beq
i\b''_{l1}rh_l+(2i\b'_{l1}-\0{\s'}{E})(h_l+rh'_l)-\0{\k\s'}{E}h_l=0,
\eeq \bea i\b''_{l2}rh_l&-&\b^{\prime
2}_{l1}rh_l-\0{i\s'}{E}\b'_{l1}rh_l+
(2i\b'_{l2}+\0{\s'\s}{E^2})(h_l+rh'_l)\nnu \\ &+&\left[\0{\k\s'\s}{rE^2}-(\s^2-\s^2_v)\right]rh_l=0,
\eea These equations could be numerically solved to obtain the
first and second Born approximations of the phase shift namely
$\d^{(1)}_l$ and $\d^{(2)}_l$ as \beq
\d^{(1)}_l=-2g\textmd{Re}\b_{l1}(k,r=0), \ \ \
\d^{(2)}_l=-2g^2\textmd{Re}\b_{l2}(k,r=0). \eeq Finally the
subtracted phase shift $\bar\d_l$ can be determined by equation
(\ref{delta}).

In Ref.\cite{ref13} it is shown that
these subtractions are added back into the energy by using their
explicit diagrammatic representation in terms of divergent
diagrams as one and two insertions of $\s$ field to the fermion
loop, which will be renormalized by the counterterms $E_{ct}$ and
yield a finite contribution denoted by $\G_2$. Thus one has the
renormalized one loop quantum correction energy \beq
E^{ren}_{vac}=-\sum\limits_{n}E_n-\sum\limits_{l}(2l+1)\int
dk\01\pi\0{d\bar\d_l(k)}{dk}E(k)+\G_2. \label{eren} \eeq

The detail calculation of the renormalized energy part $\G_2$ is in the following.
The only divergent Feynman digram which needs to be evaluated could be expressed in the following form \bea
i\Pi(q^2)=-g^2\int\0{d^4p}{(2\pi)^4}\textmd{Tr}\left[S(\slashed{p}+\slashed{q})S(\slashed{p})\right] \eea
Where $S(\slashed{p})$ is the fermion propagator. By the standard dimensional regulation one obtains the regulated result \bea
\Pi(q^2)=\0{g^2}{4\pi^2}\left\{\01{\e}(q^2+2m^2)+\016q^2+m^2\right. \nnu \\ \left.-\int_0^1dx\left[3x(1-x)q^2+m^2\right]\ln\0D{\mu^2}\right\},
\eea where $D=x(1-x)q^2+m^2$. The divergent part can be renormalized by a on-shell mass
renormalization which means \beq
\left.\Pi_{ren}(q^2)\right|_{q^2=-m^2_{\s}}=0, \ \ \ \
\left.\0{d\Pi_{ren}(q^2)}{dq^2}\right|_{q^2=-m^2_{\s}}=0, \eeq
where $m_{\s}$ is taken as the usual sigma meson mass
$m_{\s}=550MeV$. The divergent parts are removed by the counter
terms. Then the renormalized result is \bea
&&\Pi_{ren}(q^2)=\nnu \\ &&-\0{g^2}{4\pi^2}\left\{\int_0^1dx\left[3x(1-x)q^2+m^2\right]\ln\0{m^2+x(1-x)q^2}{m^2-x(1-x)m^2_{\s}}\right.
\nnu \\ &&\left.+(q^2+m^2_\s)\int_0^1dxx(1-x)\0{3x(1-x)m^2_{\s}-m^2}{m^2-x(1-x)m^2_{\s}}\right\}.
\label{pi} \eea In the following calculation I will change the
four momentum to the three momentum as $q=|\vec{q}|$ by setting
$q_0=0$. Now the finite energy term $\G_2$ can be evaluated as
\beq
\G_2=\int_0^{\infty}\0{q^2dq}{2\pi^2}\Pi_{ren}(q^2)\tilde{\s}(q)^2,
\label{gama} \eeq where $\tilde{\s}(q)$ is the Fourier transform
of $\s(\vec{r})$ which result is \beq \tilde{\s}(q)=\0{4\pi^2
r_0\s_0}{q\sinh(\pi qr_0)}\left[R\cos(qR)-\0{\pi r_0}{\tanh(\pi
qr_0)}\sin(qR)\right]. \eeq  Notice
that the homogeneous background field $\s_v$ is subtracted from
$\s(r)$ because it will generate the energy from the scattering
effect to the homogeneous vacuum background which is infinite and
not relevant to our physical result. Substituting the above result of
$\tilde{\s}(q)$ into equation (\ref{gama}) together with the
result of $\Pi_{ren}(q)$ from equation (\ref{pi}), the
renormalized energy part $\G_2$ can be numerically calculated.

\begin{figure}[tbh]
\includegraphics[width=190pt,height=130pt]{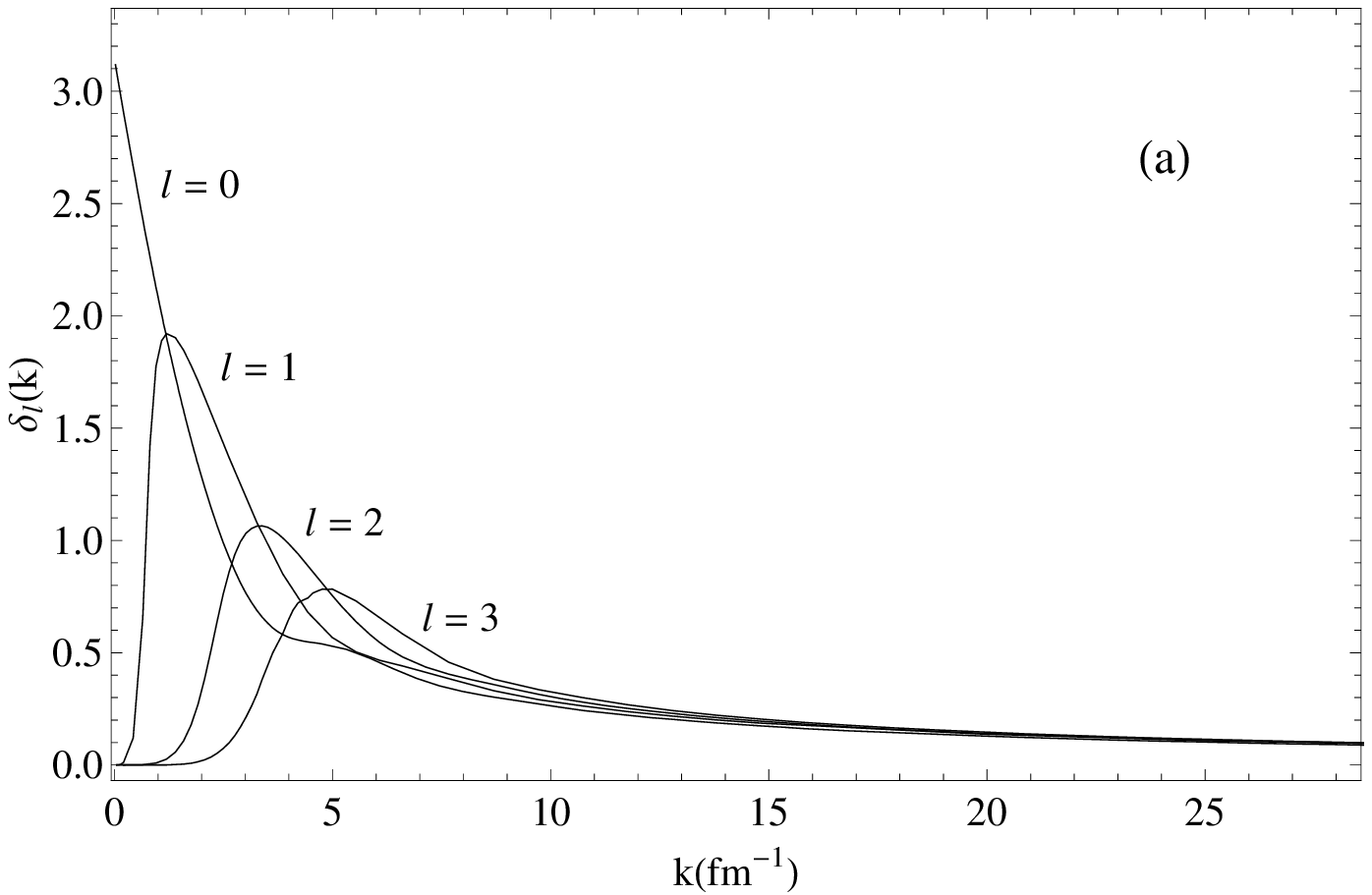}
\includegraphics[width=190pt,height=130pt]{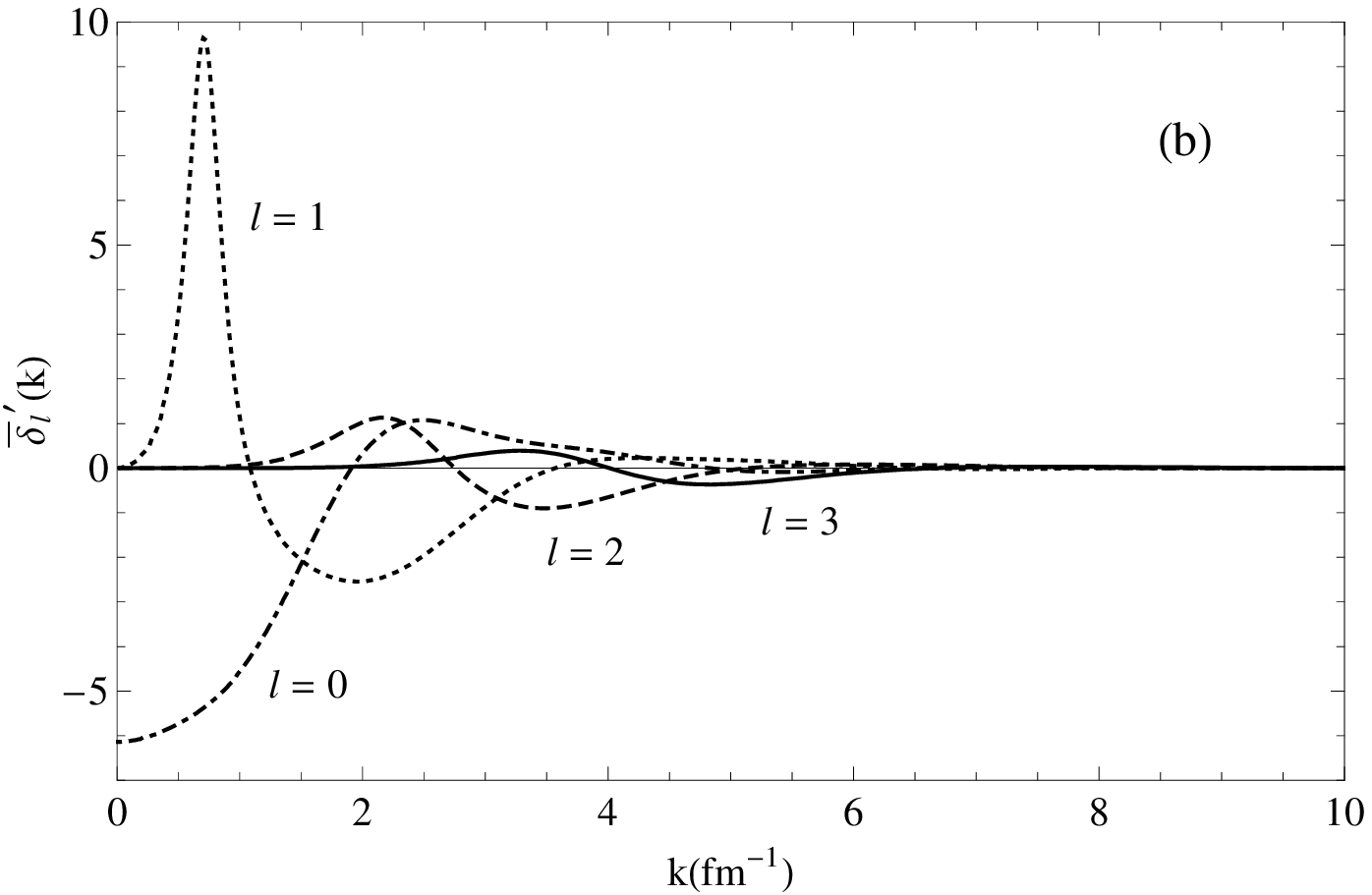}
\caption{(a)The phase shift $\d_l(k)$ without subtraction as a
function of momentum $k$, for $l=0,1,2,3$. (b)The first order momentum derivative of subtracted phase
shift $\bar\d_l(k)$ as a function of momentum $k$. The dot-dashed,
dotted, dashed and solid lines are for the cases of $l=0,1,2,3$
respectively.}\label{x}
\end{figure}

The numerical result of the phase shift $\d_l(k)$ is presented in
figure \ref{x}.(a).
It could be seen that the amplitudes of the phase shift decrease
with $l$ increasing. However at $k\to \infty$ the phase shift
decreases with momentum increasing in a logarithmic way and
approaches zero very slowly. By subtracting the Born approximation
$\d^{(1)}_l$ and $\d^{(2)}_l$ one obtain the subtracted phase
shift $\bar\d_l(k)$. In figure \ref{x}.(b) the
numerical result of $\bar\d'_l(k)$ is presented. When $k\to \infty$ they all
approach zero exponentially which makes the integral over momentum
finite in equation (\ref{eren}).
Here the energy term associated with different angular momentum
$l$ can be defined as \beq E^{(l)}=\int
dk\01\pi\0{d\bar\d_l(k)}{dk}E(k). \eeq The numerical results of
them are in the following \bea E^{(0)}=-4.26fm^{-1}, \ \
E^{(1)}=-0.53fm^{-1}, \nnu \\ E^{(2)}=-0.3fm^{-1}, \ \
E^{(3)}=-0.19fm^{-1}, \ \ ... \eea  Additionally there is only one
bound state which energy is $E_1=1.6fm^{-1}$. However the energy
level of this bound state has already been occupied by the three
valence quarks, so this bound state does not contribute to the correction energy. The vacuum correction only comes from
summing the continuum energy spectrum of the scattering states.
Considering the summation of the energy over $l$ from $l=0$ to
$l=3$ together with $\G_2$ which is $\G_2=-0.42fm^{-1}$ the renormalized quantum
correction energy will be $E^{ren}_{vac}\sim 8.26fm^{-1}$ which
magnitude has already exceeded the value of the classical energy
which is $E_{cl}\approx 6.4fm^{-1}$, not to mention the energy terms with
$l>3$. Thus the one loop quantum correction is quite large and
will not support the existence of the soliton in FL model in this
context.

In this paper I have used a practical method to study the quantum fluctuations of the non-topological solitons in FL model. One could see that it is efficient and all the divergences have been removed by the same renormalization procedure. At the present level I just focus the study at the zero temperature case. However an interesting issue is the quantum correction of soliton at finite temperature. In that case one can even study the quantum fluctuations of solitons during the deconfinement phase transiton in FL model. These calculations are doable and will be a separate work deserving thorough discussion in the next step. Another interesting issue is that the practical method could be further extended to the chiral soliton model and other soliton models based on QCD theory. In recent years the topological solitons in QCD, like instantons and dyons, have been actively studied by Shuryak and Zahed~\cite{ref18,ref19}. These nontrivial vacuum structures of QCD are believed to be able to produce both confinement and chiral symmetry breaking. Most of the calculations on instantons and dyons are semi-classical. The quantum fluctuations are also important in these systems, especially during the phase transition. However the calculations of the quantum fluctuations of the QCD dyon ensemble at finite temperature will be a very challenging task. All these issues are under consideration and will be studied in the future work.

\begin{acknowledgments}
The author S. Shu is very grateful to Edward Shuryak for his
helpful suggestions and many useful discussions. S. Shu is also
thankful to the nuclear theory group at the Stony Brook University
for their hospitality and support. This work is supported in part
by the China Scholarship Council.
\end{acknowledgments}

\end{document}